\documentclass[a4paper,12pt,reqno]{amsart}

\usepackage[centertags]{amsmath}
\usepackage{amsfonts}
\usepackage{euscript}
\usepackage{amssymb}
\usepackage{amsthm}
\usepackage{newlfont}
\usepackage{stmaryrd}
\usepackage{mathrsfs}

\theoremstyle{plain}

\theoremstyle{definition}

\theoremstyle{remark}

\numberwithin{equation}{section}

 \let\be=\beta  \let\ep=\epsilon
  \let\ga=\gamma 
 \let\la=\lambda  
\let\si=\sigma

\newcommand{\caA}{{\mathcal A}}

\newcommand{\caD}{{\mathcal D}}
\newcommand{\caE}{{\mathcal E}}

\newcommand{\caL}{{\mathcal L}}

\newcommand{\caP}{{\mathcal P}}
\newcommand{\caQ}{{\mathcal Q}}

\newcommand{\caS}{{\mathcal S}}
\newcommand{\caT}{{\mathcal T}}

\newcommand{\bbL}{{\mathbb L}}

\newcommand{\opunit}{\text{1}\kern-0.22em\text{l}}

\newcommand{\frj}{{\mathfrak j}}

\newcommand{\bsP}{{\boldsymbol P}}

\DeclareMathAlphabet{\mathpzc}{OT1}{pzc}{m}{it}

\newcommand{\rel}{\,|\,}

\newcommand{\id}{\textrm{d}}

\newcommand{\inn}{_\text{int}}
\newcommand{\ext}{_\text{ext}}

\usepackage{epsfig,amsmath,amsfonts,amsbsy}

\begin{document}

\begin{center}
\noindent {\Large {\bf On and beyond entropy production:\\ the
case of Markov jump processes}}

\vspace{15pt} {\bf Christian Maes}\footnote{Instituut voor
Theoretische Fysica, K.~U.~Leuven, Belgium.\\ {\tt
http://itf.fys.kuleuven.be/\~{}christ}},  {\bf Karel
Neto\v{c}n\'{y}}\footnote{Institute of Physics AS CR, Prague, Czech
Republic.\\ email: {\tt netocny@fzu.cz}}, and {\bf Bram
Wynants}\footnote{Instituut voor Theoretische Fysica, K.~U.~Leuven,
Belgium.\\ email: {\tt bram.wynants@fys.kuleuven.be}} \vspace{5mm}
\end{center}

\vspace{20pt} \footnotesize \noindent {\bf Abstract: } How is it
that entropy derivatives almost in their own are characterizing
the state of a system close to equilibrium, and what happens
further away from it? We explain within the framework of Markov
jump processes why fluctuation theory can be based on
considerations involving entropy production alone when perturbing
around the detailed balance condition.  Variational principles
such as that of minimum entropy production are understood in that
way. Yet, further away from equilibrium, dynamical fluctuations
reveal a structure where the time-symmetric sector crucially
enters. The fluctuations of densities and currents get coupled and
a time-symmetric notion of dynamical activity becomes the
counterpart and equal player to the entropy production. The
results are summarized in an extended Onsager-Machlup Lagrangian,
which in its quadratic approximation is expected to be quite
general in governing the small fluctuations of nonequilibrium
systems whose macroscopic behavior can be written in terms of a
Master equation autonomously describing the time-dependence of
densities and currents.
\normalsize

\vspace{10mm}
\section{Scope}
The breaking of time-reversal symmetry is certainly an important
feature of nonequilibrium systems.  While the underlying microscopic
dynamics is (under usual circumstances) time-reversal symmetric, the
plausibility of the time-reversed history of mesoscopic or even more
macroscopic conditions can greatly differ from that of the original
history. These considerations are very much linked with the concept
of entropy and its production.  As written by Max Planck in 1926
\cite{Planck}: {\it "...there is no other general measure for the
irreversibility of a process than the amount of increase of
entropy."}  As an example, the by now well-known fluctuation symmetries
of the entropy production, be it transient or in the steady state,
are on a formal level nothing but expressions of that relation
between entropy production and time-reversal breaking.  That point
was especially emphasized in
 \cite{crooks,mn,poincare}.\\

In particular and  even if not always explicit, much emphasis in
the study of nonequilibrium phenomena has gone to the study of the
entropy production, or to some nonequilibrium extension and
generalization of thermodynamic potentials. Nevertheless there are
reasons to doubt the unique relevance of the entropy production
concept, as traditionally understood, in far from equilibrium
set-ups. Similar thoughts have already been expressed longer time
ago in \cite{landauer}.
 The characterization of nonequilibrium could very well require to
consider observations that are somewhat foreign to equilibrium
thermodynamics.  Entropy production governs
 equilibrium fluctuations and remains useful for close-to-equilibrium
processes via the concept of heterogeneous equilibrium.  Yet, that
{\it hydrodynamic experience} is mostly related to the problem of
return to equilibrium. For all we can imagine, perhaps other
quantities must complement the entropy production to account for
other relevant nonequilibrium features that have to do not only
with dissipation but perhaps also with more constructive aspects
of the nonequilibrium kinetics and its
dynamical activity.\\

We are interested here in the fluctuation functionals for the nonequilibrium
statistics of state-occupations and of state-transitions. Much research into this has already been done, and for a review of this we refer to \cite{lasinio}. Our emphasis lies on the \emph{joint} fluctuations of these time-symmetric and
time-antisymmetric sectors of the dynamical fluctuations. The most important observations will be, that these sectors are coupled and that they are not
solely determined by the stationary entropy production. A similar emphasis was put
already in the treatment of the steady state statistics of
diffusions, \cite{MNW}, and in the elucidation of a canonical
structure of the steady fluctuations \cite{prl}.  Here we are
adding the discussion on the transient regime and for the steady
state we concentrate fully on the small fluctuations around the
stationary values.  That suffices to appreciate the appearance of
a new quantity, that we have called {\it traffic} and that
measures the dynamical activity in a time-symmetric way.\\
The type of nonequilibrium systems to which we believe our
analysis applies almost literally are composed of weakly
interacting particles, as in a driven dilute gas for which a
Boltzmann-Grad limit can be taken, or as in a driven Lorentz gas,
or consist of a multi-level system in contact with particle or
heat reservoirs.  The latter are frequently encountered in quantum
transport systems on the nanoscale.  When dealing with interacting
particles, we would think that the line of analysis can be kept
but interesting new behavior, including phase transitions, can
result and be accompanied by a less trivial application of the
theory of large deviations.  We refer to \cite{der} and references therein
for an update.\\
The text is not a fully mathematical treatment.  Our excuse is
that we think more important today to put the physical
concepts in place and to suggest a fruitful line of physical
reasoning. Moreover, the mathematical formulation and proofs are
expected to be rather straightforward, and not adding
substantially to the interest of the paper.  Nevertheless, we
realize that the paper can still appear a bit heavy on the formal
side; no standard examples
are included from the recent nonequilibrium literature.
  We hope that future contributions will remedy that. \\

The next section presents the set-up; the particular framework is
that of Markov jump processes to be used as start for a
nonequilibrium thermodynamics of free particles.  Section \ref{ep}
reminds us of the notion of entropy production.  A separate
section \ref{dyna} is devoted to the notion of dynamical
fluctuations. The
rest of the paper analyzes the resulting generalized
Onsager-Machlup Lagrangian, first around equilibrium in Section
\ref{eq}  and then for more arbitrary nonequilibrium conditions in
Section \ref{noneq}.

\section{Set-up}\label{sec: finite space}

Imagine a large number $N$ of degrees of freedom $
(x^1_t,\ldots,x^N_t)$ evolving in continuous time $t$.  The case
we consider is that of a collection of jump processes, with a
common state space $\Omega$.  On the level of a macroscopic
description, two types of empirical averages present themselves:
first,
\begin{equation}\label{occ}
  p^{N}_t(x) = \frac{1}{N}\sum_{k=1}^N \chi[x_t^k = x],
  \qquad x \in \Omega
\end{equation}
where the indicator function $\chi$, possibly understood in a
distributional sense, gives the state-occupation. Secondly, there
is the empirical distribution of jumps $x \rightarrow y$ for all
pairs of states $x, y \in \Omega$, that we write in the form
\begin{equation}\label{rates}
  \frac{1}{N} \#\{\text{jumps $x \rightarrow y$ within $[t,t+\tau]$}\}
  = \tau p^N_t(x) k^N_t(x,y) + o(\tau)
\end{equation}
where we have already expressed the assumption that these suffice
for an autonomous description.  In other words, the $p^N$ and
$k^N$ are not completely arbitrary but they have to satisfy the
consistency (or balance) relation
\begin{equation}\label{eq: balance}
  \frac{\id p^N_t(x)}{\id t} = \sum_{y \neq x}
  \{p^N_t(y) k^N_t(y,x) - p^N_t(x) k^N_t(x,y)\}
\end{equation}

One recognizes in \eqref{eq: balance} the form of a Master equation.
It is one of the challenges of nonequilibrium statistical mechanics
to actually obtain useful conditions on the derivation of such
Master equations from more microscopic evolution laws.  Here we
ignore that problem and we actually {\it start} from the $x^i_t$ as
a collection of independent and identical Markov jump processes.  To
make it even simpler, we assume that $\Omega$ is finite and that the
process $(x_t^i)$ is ergodic with unique stationary law $\rho$. All
that is believed not to be extremely important, as we have in mind
the $N\uparrow +\infty$ limit. The possibility of phase-transitions or of {\it
non-smooth} behavior is not considered here in exchange for a
thorough look at the fluctuation theory of the $(p^N_t,k^N_t)_t$ and
derived quantities.

\subsection{Macroscopic limit}

We consider a collection of identical independent ergodic continuous
time Markov processes $x^i_t$, each taking values in the finite
state space
$\Omega$ and with rates $\lambda(x,y)\geq 0$ for jumps between the
states $x\rightarrow y$. We interpret the process
$(x^i_t)_t$ as the random trajectory of the $i$-th particle, where
randomness refers to some reduced description where further degrees
of freedom are integrated out possibly in combination with some
particular limiting procedure.   On the macroscopic level, we deal
with the trajectory
$(p^N_t,k^N_t)_t$ from \eqref{occ} and \eqref{rates}. It defines the
whole empirical process which is (time-inhomogeneous) Markov even
for finite $N$ by construction; note that we do not include three-
and higher-time empirical correlations into our macroscopic
description. From the law of large numbers, the random occupations
$(p^N_t)_t$ concentrate in the limit $N\to\infty$ on the unique
solution of the Master equation
\begin{equation}\label{timev}
  \frac{\id p_t(x)}{\id t} =
   \sum_{y \neq x} \{p_t(y) \la(y,x) - p_t(x) \la(x,y)\}
\end{equation}

\subsection{Path distribution}
The trajectories of the particle do not all have the same
probability. And the same trajectory has different probabilities
depending on the rates of the process.   All that can be studied
via standard tools for comparing probability densities, in
particular via the so called Girsanov formula for Markov
processes. For our context,
 the density of one path-space
measure $\cal{P}_{\mu}$ over a time $T$, starting at probability
law $\mu$, with respect to another one $\bar{\cal{P}}_{\bar{\mu}}$
is given by
\begin{equation}\label{girs}
\frac{\id\caP_{\mu}}{\id\bar \caP_{\bar{\mu}}}(\omega) =
\frac{\mu(x_0)}{\bar \mu(x_0)}\exp \Bigl\{ -\int_0^T \,
\bigl(\xi(x_t) - \bar \xi(x_t) \bigl)\,\id t  + \sum_{0<t< T}\log
\frac{\lambda(x_{t^-},x_{t})}{{\bar \lambda}(x_{t^-},x_{t})}
\Bigr\}
\end{equation}
where $\omega= (x_t)_0^T, x_t\in \Omega,$ is a piecewise constant
trajectory (or path) with, as first integral in the exponent
\[ \xi(x) = \sum_{y \neq x} \lambda(x,y)\]
the escape rates, and with the last sum in the exponent being over
the jump times $t$ where the path takes $x_t$ to $x_{t^+}$. As a
convention, we always take right-continuous versions of the
trajectories. As usual with probability densities, there is the assumption of absolute continuity making the
undefiniteness not worse than giving weight zero to terms of the
form $0/0$. Mathematical details and
derivation can be found in e.g. Appendix~2 of~\cite{KL}.\\
That is useful for our fluctuation theory as we can obtain the
probability of an event as the density of the original process with
respect to a new process which makes the event typical, conditioned
on that event. That is sometimes referred to as the Cramer-trick in
the theory of large deviations;
a gentle introduction is contained in \cite{var}.\\

Remark that the exponent in \eqref{girs} contains two terms, the
first one (with the escape rate) is time-symmetric, the second one
is time-antisymmetric.  In fact, soon we will see (in Sections
\ref{si} and \ref{fr}) that the time-antisymmetric part in the
action governing the path-space distribution is exactly the
entropy production.

\subsection{Relation to thermodynamics: local detailed balance}
Up to here, we have only statistically defined our model. To get a physical (measurable) interpretation we should associate thermodynamics to it.
In the case of equilibrium, this is well-known: by equilibrium we mean that case where
\begin{equation}\label{db}
\rho(x)\, \lambda(x,y) = \rho(y)\,\lambda(y,x)
\end{equation}
where $\rho(x) \propto e^{-\beta U(x)}$ is a Gibbs-distribution.
This relation expresses a reversal symmetry for each of the transitions
$x\leftrightarrows y$, which finally amounts to the time-reversal
symmetry of the stationary process.  We restrict us here to state
spaces for which the kinematical time-reversal is trivial (no
velocities).  That is a serious restriction, which is typical for
chemical reaction networks or for overdamped motion but one should understand that it
greatly influences the relation between time-reversal, equilibrium and entropy production. \\

For models of nonequilibrium systems, a thermodynamic interpretation becomes difficult because one expects that the condition
of detailed balance \eqref{db} is broken. What replaces it, is
either derived from more microscopic models or is assumed. What
guides that procedure is known as the condition of local detailed
balance.  For our purposes we can write it in terms of an energy
function $U(x)$ and a work function (or driving) $F(x,y) = -F(y,x)$
to assign rates to the transitions between each  $x$ and $y$,
satisfying
\begin{equation}\label{ldb}
\frac{\lambda(x,y)}{\lambda(y,x)}=e^{\beta(F(x,y)+U(x)-U(y))}
\end{equation}
where $\beta\geq 0$ is a parameter that stands for the inverse
temperature of a reference reservoir.\\
The fundamental reason for local detailed balance is the
time-reversibility of an underlying microscopic dynamics over which
our effective stochastic model is presumably built. Hence, violating
such a condition reduces the physical interpretation of our
stochastic model. As further explained in the next section, condition \eqref{ldb} is
also intimately related to the symmetries of nonequilibrium
fluctuations and to the role of entropy production in there.
At any event, \eqref{ldb} allows to write
\begin{equation}\label{transrates}
\rho_o(x)\lambda(x,y) = \gamma(x,y)\,e^{\frac{\beta}{2}F(x,y)}
\end{equation}
with $\rho_o(x) = e^{-\beta U(x)}/Z$ a reference equilibrium
probability distribution and some symmetric $\gamma(x,y) =
\gamma(y,x)$, which is left unspecified. To reveal the meaning of $\gamma$, notice that in equilibrium, i.e.\
for
$\rho = \rho_o$ and $F = 0$, one has
$2\gamma(x,y) = \rho(x)\, \lambda(x,y) + \rho(y)\,\lambda(y,x)$. The
right-hand side is the expectation of the empirical observable
\begin{equation}\label{traf}
  \tau^N_t(x,y)= p_t^N(x)\, k^N_t(x,y) + p_t^N(y)\, k^N_t(y,x)
\end{equation}
that measures the time-symmetric dynamical activity (the total number of jumps across the bond $(x,y)$) and we call it
\emph{traffic}. As we will see in~Sections~\ref{noneq-full}~and~\ref{strucnor}, the traffic
is a crucial quantity to characterize the nonequilibrium
fluctuations far from equilibrium.

\section{Entropy production}\label{ep}

The notion of entropy production should not be fully re-invented
when dealing with Markov jump processes.  It must match with the
thermodynamic or hydrodynamic interpretations.  We start however
with a view that goes beyond model-specifics and that emphasizes
the relation with time-reversal.

\subsection{Statistical interpretation:
time-irreversibility}\label{si}

Dynamical time-reversal plays on the level of single trajectories
$\omega = (x_t)_0^T$. We define the time-reversal as $\theta
\omega = (x_{T-t})_0^T$, not indicating the trivial modifications at
the jump times for restoring the right-continuity of paths. If we
denote the original Markov process started at distribution $\mu$
by $\cal{P}_{\mu}$, then there is a time-reversed process
$\cal{P}_{\mu_T}\theta$ starting at the (time-evolved)
distribution $\mu_T$.  There is a density of one with respect to
the other, and that we call the (variable, fluctuating) entropy
production
\begin{equation}\label{vep} S^T_{\mu}(\omega) =
\log\frac{\id\cal{P}_{\mu}}{\id\cal{P}_{\mu_T}\theta}(\omega)
\end{equation}
 We can use the Girsanov formula (\ref{girs}) for its
computation, see the details in \cite{jmp}. That formula
\eqref{vep} captures the idea of the entropy production as
measuring the amount of time-reversal breaking.  The so called
fluctuation theorem, steady or transient, time-dependent or not,
very much rests on that unifying idea, \cite{poincare}. For a
foundation starting from the Hamiltonian dynamics and microcanonical
ensemble, see
\cite{mn}.  We come back to fluctuation relations in Section
 \ref{fr}.\\
It is interesting to note that by convexity
\[
\left< S^T_\mu \right>_{\mu} \geq 0
\]
as it should for an entropy production,
 where the brackets take the average with respect to
$\cal{P}_{\mu}$, the path-space measure starting at $\mu$.

Being interested in an instantaneous (average) entropy production
rate when the distribution is $\mu$, we define
\begin{equation}
\sigma(\mu) = \lim_{T \downarrow 0}\frac{1}{T}
\left<\log
\frac{\id\cal{P}_{\mu}}{\id\cal{P}_{\mu_T}\theta}
\right>_{\mu}\label{alent}
\end{equation}
so that, by the Markov property,
$\left< S^T_\mu \right>_{\mu} = \int_0^T \sigma(\mu_t)\,\id t$ with
$(\mu_t)_0^T$ the time-evolved measures. The instantaneous entropy production
rate
$\sigma(\mu)$ can easily be computed for our Markov process:
\begin{equation}\label{entprod}
\sigma(\mu) = \sum_{(xy)}[\mu(x)\lambda(x,y) -
\mu(y)\lambda(y,x)]\log\left(\frac{\mu(x)\lambda(x,y)}
{\mu(y)\lambda(y,x)}\right)
\end{equation}
The notation $(xy)$ under the sum will from now on be used to mean
that we sum over unordered pairs of states. \\
The previous expressions do make physical sense even for a single
Markov process defining a dynamics for a small finite number of
degrees of freedom, thinking of an open system effectively coupled
to and/or driven by large external reservoirs. It becomes however
more physically transparent when formulated in terms of the
empirical distribution as explained next.

\subsection{Thermodynamic interpretation}
An open system dissipates heat that results in
  a change of entropy in the environment.
Assuming a large environment we can compute it as the reversible
heat.   From the first law of thermodynamics that dissipated heat
is identical to the  work plus the change in internal energy. So,
again in our ensemble-interpretation, the rate of change of energy
is $- \sum_{x,y} j^N_t(x,y)\,U(x)$ with, see \eqref{eq: balance},
\begin{equation}\label{empcur}
j^N_t(x,y) =  p^N_t(x) k^N_t(x,y) - p^N_t(y) k^N_t(y,x)
\end{equation}
being the empirical currents, and the power is
$\sum_{x,y} j^N_t(x,y)\, F(x,y)$.  If therefore the empirical currents at time $t$ equal
$j^N_t(x,y) = j(x,y)$, then the dissipated heat is
\begin{equation}
  \caQ(j) = \sum_{(xy)}j(x,y)\bigl(U(x) - U(y) + F(x,y)\bigr)
\end{equation}
and the entropy current is $\beta   \caQ(j)$
 (setting
Boltzmann's constant equal to one) for an environment at
temperature $\beta^{-1}$.\\
Secondly, there is the change of the entropy of the system itself.
Here we only have the densities \eqref{occ} as macroscopic
variable and
\begin{equation}\label{static-ent}
  S_{\text{sys}}(p) = -\sum_{x} p(x) \log p(x)
\end{equation}
is the (static) fluctuation functional in the probability law for
observing the empirical density $p$ when sampling the particles from
the flat distribution.  Its change in time is the internal entropy
production:
\[
\dot S_{\text{sys}}(p,j) = \sum_{(xy)}j(x,y)(\log p(x) -
\log p(y))
\]
Summing it up, we get the total (macroscopic) entropy production
rate
\begin{equation}\label{macroepi}
  \dot{S}(p,j) \equiv \dot S_{\text{sys}}(p,j) + \beta\caQ(j)
\end{equation}
for the empirical values $p$ and $j$ for, respectively densities
\eqref{occ} and currents \eqref{empcur}.

\subsection{Relating the two interpretations}
Using the local detailed balance condition (\ref{ldb}), the
macroscopic entropy production rate in \eqref{macroepi} is
\begin{equation}\label{macroep}
\dot{S}(p,j) = \sum_{(xy)}j(x,y)
\,\log \frac{p(x)\lambda(x,y)}{p(y)\lambda(y,x)}
\end{equation}
in terms of the instantaneous densities $p(x)$ and currents
\[
j(x,y) = p(x)k(x,y) - p(y)k(y,x)
\]
Remember that $p(x)k(x,y)$ is the fraction of particles that
actually make the transition $x\to y$, and $j(x,y)$ is the (net)
current of particles. By the law of large numbers, the typical value
of these currents at given densities $p(x)$ is
$p(x)\la(x,y) - p(y)\la(y,x)$ and hence the typical entropy
production rate~\eqref{macroep} just coincides with
$\si(p)$, see~\eqref{entprod}. This not only justifies our form of the local detailed balance assumption, it also explains the relation between
the single Markov process formalism of Section~\ref{si} and the
empirical description for an ensemble of the processes,
cf.~\eqref{occ}--\eqref{rates}; this duality is exploited throughout
the whole text.\\

A different decomposition that is equally useful (and used in the
following subsection) writes
\begin{equation}\label{dec}
\dot{S}(p,j) =  \dot S\ext(j) + \dot{S}\inn(p,j)
\end{equation}
for the entropy current
\begin{equation}\label{ent-current}
  \dot S\ext(j) = \sum_{(xy)} j(x,y)\,F(x,y)
\end{equation}
in excess with respect to the equilibrium
reference, and
\[
  \dot{S}\inn(p,j) = \sum_{(xy)} j(x,y)
  \Bigl( \log \frac{p(x)}{\rho_o(x)} - \log \frac{p(y)}{\rho_o(y)}
  \Bigr)
\]
is now the rate of change of the system's entropy (always summing
over pairs). Note that this decomposition differs only from the former (\ref{macroepi}) in the use of another reference. In (\ref{macroepi}) the reference is the flat distribution. To end this section we review two simple applications
of the single process formalism of Section~\ref{si}.

\subsection{Fluctuation relations}\label{fr}

The decomposition~\eqref{dec} into the internal and external change
of the entropy can equivalently be done pathwise for a single process,
starting from~\eqref{vep}. Note that it depends on the choice of the
reference equilibrium process; using the notation $\caP_o$ for such a
reference started from a reversible measure
$\rho_o$. Then~\eqref{girs} can be written in the form
\[
\id\cal{P}_\mu(\omega) =
\id\cal{P}_o(\omega)\,\frac{\mu(x_0)}{\rho_o(x_0)}\,e^{-A(\omega)}
\]
with the action $A$ that can be read from \eqref{girs}.  Since the
reference process is time-reversal invariant, we can now rewrite
\eqref{vep} as
\[
S_\mu^T(\omega) =
\log\frac{\mu(x_0)\,\rho_o(x_T)}{\mu_T(x_T)\,\rho_o(x_0)}\,+
A(\theta\omega) - A(\omega)\] That corresponds to the decomposition
\eqref{dec} and we call
\[
S\ext(\omega) = A(\theta\omega) - A(\omega)\] the (variable) entropy
flux for a single chain, in excess with respect to the reference
equilibrium process.  Obviously, we also have
\[
S\ext(\omega)  =
\log\frac{\id\cal{P}_{\rho_o}}{\id\cal{P}_{\rho_o}\,\theta}(\omega)
\] and hence, for all path-dependent observables $f$,
\begin{equation}\label{fluc}
  \langle f\rangle_{\rho_o} = \langle f\theta\;
  \exp(-S\ext)\rangle_{\rho_o}
\end{equation}
which gives an exact (for all finite times $T$) symmetry in the
distribution of the (excess) entropy flux
$S\ext = -S\ext\theta$ at least when started from equilibrium.
Steady fluctuation symmetries are then
obtained as the asymptotics for $T\uparrow +\infty$. Note that in
general one needs to deal with the temporal boundary term. However,
in the present framework of ergodic Markov processes over a finite
state space the dependence on the initial condition is irrelevant.\\
The fluctuation symmetry~\eqref{fluc} also has a formulation in
terms of the macroscopic fluctuation theory within the ensemble
formalism of the previous subsection; this will be discussed at the
end of Section~\ref{sec: lagrangian-currents}.

\subsection{Stationary measure}

One may wonder how the above considerations are reflected on the
level of the stationary distribution $\rho$ itself.  That in fact is
the subject of earlier  work by Zubarev and by MacLennan,
\cite{McL}: what is a first order correction around
a reference equilibrium/detailed balance, and is there a systematic
perturbation theory? There are a number of ways to discuss that
question. One possible direction is to try to formulate a
variational principle for the $\rho$; this approach will be
discussed later. Another, more direct approach is to compute the
asymptotics $T\uparrow+\infty$ of the time-evolved measure $\mu_T$
or, equivalently, to project the path-space distribution
$\caP^T_\mu$ on the time $T$, again asymptotically. As explained in a recent preprint,
\cite{kn}, the latter approach can be conveniently started from the fluctuation
symmetry~\eqref{fluc}. Indeed, by taking $f(\omega) = \chi[x_T=x]\,\exp(-S\ext(\omega)/2)$, one has
$f\theta(\omega)= \chi[x_0=x]\,\exp(S\ext(\omega)/2)$ and therefore
\[
  \Bigl\langle \chi[x_T=x]
  \exp\Bigl(-\frac{S\ext}{2}\Bigr) \Bigl\rangle_{\rho_o}
  = \Bigl\langle \chi[x_0=x]\,\exp\Bigl(-\frac{S\ext}{2}\Bigr) \Bigr\rangle_{\rho_o}
\]
As a consequence, the probability to see $x$ at time $T$ when
started from reference equilibrium $\rho_o$ is
\begin{equation}\label{mcl-kn}
  \bsP_{\rho_o}^T(x_T = x) = \rho_o(x)\,\frac{\bigl\langle
  \exp\bigl(-\frac{S\ext}{2}\bigr) \bigr\rangle_{x_0=x}}{\bigl\langle
  \exp\bigl(-\frac{S\ext}{2}\bigr) \bigr\rangle_{x_T=x,\,\rho_o}}
\end{equation}
where we have to condition on the final-time event $x_T=x$ in the
denominator. The ratio is one for the equilibrium dynamics, and the
nonequilibrium correction is made by the time-asymmetry in the
fluctuations of the entropy production. An advantage of this
representation of the evolved measure lies in the cancelation of
various nontransient (i.e.\ unbounded upon $T$ growing) terms when
expanding the exponents, so that the limit $T\uparrow +\infty$ can
be controlled,  see~\cite{kn} for more details.

\section{Dynamical fluctuations}\label{dyna}

In the ensemble picture, trajectories
$\omega^N = (x_t^1,\ldots,x_t^N)_t$ have their coarse-grained counterparts in
the empirical distributions $(p_t^N)_t$ and the empirical rates
$(k_t^N)_t$.
They fluctuate around their typical values $\rho$ and $\lambda$, the
typicality being in the sense of a law of large numbers with $N$ as the
large parameter. Computing the probability of the event
$p^N_t = p_t$, $k^N_t = k_t$ for all $0 \leq t \leq T$
is done via the Girsanov formula~\eqref{girs}, by comparing the
system with modified dynamics (such that $p_t$ and $k_t$ are
typical) with the original system. Clearly, a macroscopic trajectory
$(p_t,k_t)_t$ satisfying the consistency condition~\eqref{eq: balance}
becomes typical under the modified (time-inhomogeneous) Markov
dynamics with the rates $(k_t)_t$ and the initial measure $p_0$. Its
probability with respect to the original i.i.d.\ Markov processes
$x^1,\ldots,x^N$ started each from the distribution $\mu$ has the
large deviation form
\begin{equation}\label{eq: k-LD}
  \bsP^N_\mu\{(p^N_t = p_t,k^N_t = k_t)_{0 \leq t \leq T}\}
   \doteq \exp \Bigl\{-N\Bigl[S(p_0 \rel \mu)
  + \int_0^T \id t\,\bbL(p_t,k_t)\Bigr]\Bigr\}
\end{equation}
where $\doteq$ refers to the logarithmic equivalence as $N \to
\infty$, and the relative entropy $\caS$ and the Lagrangian $\bbL$
are
\begin{align}
  \caS(p_0 \rel \mu) &= \sum_x p_0(x) \log\frac{p_0(x)}{\mu(x)}
\\
  \bbL(p,k) &= \sum_{x, y \neq x} p(x) \Bigl[ k(x,y) \log \frac{k(x,y)}{\la(x,y)} - k(x,y) +
  \la(x,y) \Bigr]
\end{align}
In particular, the Lagrangian on this level of description has a
simple explicit form, irrespective of any detailed balance or
stationarity assumptions. It is therefore a natural point of
departure for the investigation of also more coarse-grained
dynamical fluctuations. Next we consider one step of such a conceivable hierarchy.

\subsection{Lagrangian for currents}\label{sec: lagrangian-currents}

A quantity of special interest is the collection of empirical
currents $j_t^N(x,y)= -j_t^N(y,x)$ given in \eqref{empcur}.
The Lagrangian $\caL(p,j)$ that governs the joint occupation-current
dynamical fluctuations is
\begin{equation}\label{eq: L2-gen}
  \caL(p,j) = \inf_k \{\bbL(p,k) \rel p(x) k(x,y) - p(y) k(y,x) = j(x,y);\,
  \forall x,y\}
\end{equation}
where $j$ is an arbitrary antisymmetric current matrix. The
distribution of empirical trajectories $(p^N_t,j^N_t)_t$ follows from the large deviation law~\eqref{eq: k-LD} via the
contraction principle:
\[
  \bsP^N_\mu\{(p^N_t = p_t,j^N_t = j_t)_{0 \leq t \leq T}\}
   \doteq \exp \Bigl\{-N\Bigl[S(p_0 \rel \mu)
  + \int_0^T \id t\,\caL(p_t,j_t)\Bigr]\Bigr\}
\]
whenever the consistency constraint~\eqref{eq: balance},
$\dot{p}_t(x) + \sum_{y\neq x} j_t(x,y) = 0$, is satisfied.
It can be made explicit by the method of Lagrange multipliers. One
finds
$\caL(p,j) = \bbL(p,k^j)$ where the $k^j$ solve the equations
\begin{align}\label{eq: kj}
  k^j(x,y) &= \la(x,y)\, e^{\frac{\beta}{2}\psi^j(x,y)}
\\ \label{eq: constr}
  j(x,y) &= p(x) k^j(x,y) - p(y) k^j(y,x)
\end{align}
for some specific $\psi^j(x,y) = -\psi^j(y,x)$, or explicitly,
\begin{equation}\label{eq: j-explicit}
  k^j(x,y) = \frac{1}{2p(x)}\{j(x,y)
  + [j^2(x,y) + 4 p(x) p(y) \la(x,y) \la(y,x)]^{\frac{1}{2}}\}
\end{equation}
Hence, the typical macroevolution constrained by fixing the
currents to some $j$ becomes \emph{unrestrainedly typical} by
modifying correspondingly the antisymmetric part of the transition
rates, as in (\ref{eq: kj}).\\

From (\ref{eq: j-explicit}) we check that
$p(x)k^j(x,y)=p(y)k^{-j}(y,x)$. It is then straightforward to
derive that
\begin{equation}
\caL(p,-j)-\caL(p,j) = \dot{S}(p,j)
\end{equation}
so that the entropy production rate \eqref{macroep} is (indeed)
the time-antisymmetric part of the Lagrangian.  As a consequence,
always in the logarithmic sense and in the limit $N\uparrow
+\infty$,
\begin{multline}
  \frac{\bsP^N_\mu\{(p^N_t = p_t,j^N_t = j_t)_{0 \leq t \leq T}\}}
  {\bsP^N_{\mu_T}\{(p^N_t = p_{T-t},j^N_t = -j_{T-t})_{0 \leq t \leq T}\}}
\\
  \doteq \exp \Bigl\{N\Bigl[S(p_T \rel \mu_T)-S(p_0 \rel
  \mu)\bigr] + \int_0^T \id t\,\dot{S}(p_t,j_t)\Bigr]\Bigr\}
\end{multline}
which is a macroscopic variant of the fluctuation relations
described in~Section~\ref{fr}.

\subsection{Fluctuations of empirical time averages}\label{fluemp}

Instead of looking at the probabilities of
 specific macroscopic trajectories of the system, we now
  consider empirical time averages:
\begin{equation}\label{timeav}
  \bar p_T^N = \frac{1}{T} \int_0^T p_t^N\,\id t,\qquad
  \bar j_T^N = \frac{1}{T} \int_0^T j_t^N\,\id t
\end{equation}
For a fixed initial distribution $\mu$, the asymptotic ($T\uparrow
+\infty$) probability that the empirical time averages are equal to
some density $p$ and current $j$ is
\begin{equation}
  \bsP^{N,T}_\mu\{\bar p_T^N = p,\bar j_T^N = j\} \doteq
  e^{-N \caA_T(p,j)}
\end{equation}
with the rate $\caA_T$ given by
\begin{equation}\label{eq: emp-average-action}
  \caA_T(p,j) = \inf_{p_t,j_t} \Bigl\{S(p_0 \rel \mu)
  + \int_0^T \id t\,\caL(p_t,j_t)\, \Bigl|\, \bar p_T = p, \bar j_T = j\Bigr\}
\end{equation}
where the infimum is over all macrotrajectories
$(p_t,j_t)_{0 \leq t \leq T}$ such that
$\dot{p}_t(x) + \sum_{y\neq x} j_t(x,y) = 0$. The infimum in~\eqref{eq: emp-average-action}
is easy to compute in the limit of infinite time span,
$T\uparrow +\infty$, in which the minimizing trajectory becomes essentially constant,
$(p_t, j_t)_t \equiv (p,j)$, and the initial distribution loses its relevance. One obtains
$\caA_T(p,j) = T\,\caL(p,j) + o(T)$, yielding
\begin{equation}\label{statfluct}
  \bsP^{N,T}\{\bar p_T^N = p,\bar j_T^N = j\} \doteq
  e^{-NT \caL(p,j)}
\end{equation}
whenever the currents are stationary, $\sum_{y \neq x} j(x,y) = 0$.
The equality is meant in the logarithmic sense and after taking
first the limit $N\uparrow +\infty$ and then the limit
$T\uparrow +\infty$. For details on these manipulations and techniques from the
theory of large deviations, we refer to \cite{DZ,denH,HS,DV,var}.\\

It thus appears that the Lagrangian $\caL$ of \eqref{eq:
L2-gen}--\eqref{eq: j-explicit} governs the joint steady statistics
of time-averaged occupations and currents. Its study has already
been started in \cite{prl,MNW}, emphasizing a canonical structure.
As it is fully explicit, one can use it as variational functional to
characterize the steady state, and for further contractions to
obtain variational functionals for the occupations and currents
separately.  That is however not the subject of the present paper.
What comes in the sequel is an analysis of the structure of the
above Lagrangians in the quadratic approximation and its physical
consequences.

\section{Structure of Normal fluctuations}\label{strucnor}

The most accessible fluctuations are
small---both mathematically and practically. The Lagrangians can be
expanded in both the densities and currents around their typical (=
steady) values and the strictly positive quadratic form obtained in
the leading order describes normal fluctuations. From a physical
point of view, the structure of these normal fluctuations have been
first analyzed by Onsager and Machlup, \cite{OM}, for the case of
relaxation to equilibrium. Here we show a natural extension of the
original Onsager-Machlup formalism to nonequilibrium systems by
starting from the above macroscopic fluctuation theory.

We distinguish the two following scaling regimes: first, we analyze
the nonequilibrium fluctuations in the immediate vicinity of a
reference equilibrium through the scaled Lagrangian
\begin{equation}\label{eq: scaled-0}
  L^0(u,j;F) = \lim_{\ep\downarrow 0} \ep^{-2}
  \caL(\rho_o + \ep\rho_o u, \ep j; \ep F)
\end{equation}
where we have explicitly denoted here the dependence on the work
function $F$ as it quantifies the distance from equilibrium and
participates in the scaling. The Lagrangian corresponds to the
dynamics~\eqref{transrates} with some $\ga$, $\rho_0$, and $\be$
fixed.\\
Second, we consider a steady state arbitrarily far from equilibrium
and examine the structure of small deviations through the function
\begin{equation}\label{eq: scaled}
  L(v,\frj;F) = \lim_{\ep\downarrow 0} \ep^{-2}
  \caL(\rho + \ep\rho\,v, \bar j + \ep \frj; F)
\end{equation}
Note that the work function is kept fixed here and both the density
and current are expanded around the stationary values $\rho$
respectively $\bar j$. We will see below how the structure of
fluctuations remarkably changes in this regime.

\subsection{Close to equilibrium}\label{eq}

Starting from the dynamics with transition rates parameterized as
in~\eqref{transrates},
$\la(x,y) = \rho_0^{-1}(x)\,\ga(x,y)\, e^{\frac{\beta}{2}F(x,y)}$, the scaled
Lagrangian~\eqref{eq: scaled-0} is easily computed from~\eqref{eq:
L2-gen}--\eqref{eq: j-explicit}:
\begin{equation}\label{eq: L0-explicit}
\begin{split}
  L^0(u,j;F) &= \sum_{(xy)} \frac{1}{4\ga(x,y)}
  \bigl\{j(x,y) - \ga(x,y) [u(x) - u(y) + \beta F(x,y)] \bigr\}^2
\\
  &= \frac{1}{2} \Bigl[ \frac{1}{2}\caD(j)
  + \frac{1}{2}\caE(u) - \dot s(u,j) \Bigr]
\end{split}
\end{equation}
where we have introduced the scaled entropy production,
cf.~\eqref{macroep},
\begin{equation}\label{eq: ep-linearized}
\begin{split}
  \dot s(u,j) &= \lim_{\ep\downarrow 0} \ep^{-2}
  \dot S(\rho_o + \ep\rho_o u,\ep j; \ep F)
\\
  &= \sum_{(xy)} j(x,y) [u(x) - u(y)+\beta F(x,y)]
\end{split}
\end{equation}
and the pair of (variant Onsager-Machlup) dissipation functions
\begin{align}
  \caD(j) &= \sum_{(xy)}
   \frac{j^2(x,y)}{\ga(x,y)}\label{dis1}
\\ \intertext{and}
  \caE(u) &= \sum_{(xy)}
  \ga(x,y)[u(x) - u(y)+\beta F(x,y)]^2
  \label{dis2}
\end{align}
In contrast with the equilibrium Onsager-Machlup theory, \cite{OM},
we keep here the currents $j$ as the variables of the Lagrangian.

Fixing some $u$, the typical current $j^u$ minimizes the Lagrangian,
which has the immediate solution
\begin{equation}\label{eq: typical current}
  j^u(x,y) = \ga(x,y)[u(x) - u(y) + \be F(x,y)]
\end{equation}
That variational problem for $j^u$, i.e.\
$\frac{1}{2}\caD(j) - \dot s(u,j) = \min$, is known as the
Onsager \emph{least dissipation principle}. Equivalently, it is
sometimes formulated as a \emph{transient maximum entropy production
principle}: the $j^u$ solves $\dot s(u,j) = \max$ under the
constraint
$\caD(j) = \dot s(u,j)$.

One checks that the dissipation function $\caE(u)$ is a scaled
version of the mean entropy production rate~\eqref{entprod}:
\begin{equation}\label{eq: si-E}
  \caE(u) = \lim_{\ep\downarrow 0} \ep^{-2}
  \si(\rho_o + \ep\rho_o u; \ep F)
\end{equation}
where $F$ again enters via the parametrization of the
rates~\eqref{transrates}. This is in accord with the equality
\begin{equation}\label{eq: all equal}
  \dot s(u,j^u) = \caE(u) = \caD(j^u)
\end{equation}
following from~\eqref{eq: L0-explicit}.

The consistency condition~\eqref{eq: balance} determines the typical
macroscopic trajectory as a solution of the equation
\begin{equation}
  \rho_o(x) \frac{\id u_t(x)}{\id t} + \sum_{y \neq x}
  j^{u_t}(x,y) = 0
\end{equation}
for all $x$, with $j^u$ the typical current~\eqref{eq: typical
current}. In particular, the stationary distribution
$\rho = \rho_0 + \ep \bar u + o(\ep)$ is in this
scaling limit found from the (linearized version of the) Master equation
\begin{equation}
  \sum_{y \neq x} \ga(x,y)[\bar u(x) - \bar u(y) + \be F(x,y)] = 0
\end{equation}
that has to be solved under the normalization constraint
$\sum_x \rho_o(x)\, \bar u(x) = 0$. Alternatively, the (linearized)
stationary density $\bar u$ and the corresponding steady current
$\bar j = j^{\bar u}$ can be found by minimizing the
Lagrangian~\eqref{eq: L0-explicit} subject to the stationary
constraint $\sum_{y \neq x} j(x,y) = 0$. Note a remarkable
simplification: because of this constraint, the entropy
production~\eqref{eq: ep-linearized} equals
$\dot s(u,j) = \be \sum_{(xy)} j(x,y)\, F(x,y)$ and hence it is
independent of $u$. As a consequence, the density and current become
decoupled in the Lagrangian~\eqref{eq: L0-explicit}. By the
arguments of Section~\ref{fluemp} this means that the time-averages
$\bar p_T$ and $\bar j_T$ are uncorrelated in the close-to-equilibrium regime
and within the quadratic approximation. An immediate consequence of
this observation is a simple structure of the marginal distributions
of $u$ respectively $j$ that provides a fluctuation-based
justification of the two familiar stationary variational
principles---the minimum and the maximum entropy production
principles---as we explain next. See~\cite{jmp2,BMN,prl,MNW} for
some more details and illustrations.

\subsubsection*{MinEP principle}

We consider the marginal distribution of the empirical time-average
$\bar p_T^N$ defined in~\eqref{timeav}. By~\eqref{statfluct} and
in the present scaling limit the asymptotic law of $\bar p_T^N$
reads
\begin{equation}\label{eq: occupation-law}
\begin{split}
  -\lim_{\ep\downarrow 0} \ep^{-2} \lim_{T\uparrow +\infty}
  &\lim_{N\uparrow +\infty} \log
  \bsP^{N,T}\{\bar p_T^N = \rho_o + \ep\rho_o u\}
\\
  &= \inf_j \Bigl\{L^0(u,j;F) \,\Bigl|\, \sum_{y \neq x} j(x,y) = 0\Bigr\}
\\
  &= \frac{1}{4} \bigl[ \caE(u) - \caE(\bar u) \bigr]
\end{split}
\end{equation}
where the last equality follows from~\eqref{eq: all equal} by using
the decoupling between $u$ and $j$ under the stationary constraint.
The minimum entropy production principle immediately follows:
$\caE(u) \geq \caE(\bar u)$ with the equality only if
$u = \bar u$, hence, the stationary measure minimizes the entropy
production rate, cf.~\eqref{eq: si-E}.

\subsubsection*{MaxEP principle}\label{sec: maxep}

We proceed analogously for the time-averaged empirical
current~\eqref{timeav}. For any $j$ satisfying the stationary
condition
$\sum_{x,y} j(x,y) = 0$ we have by the contraction principle
from~\eqref{statfluct}:
\begin{equation}\label{eq: current-law}
\begin{split}
  -\lim_{\ep\downarrow 0} \ep^{-2} \lim_{T\uparrow +\infty}
  &\lim_{N\uparrow +\infty} \log
  \bsP^{N,T}_\mu\{\bar j_T^N = \ep j\}
  = \inf_u L^0(u,j;F)
\\
  &= \frac{1}{2} \Bigl[ \frac{1}{2} \caD(j) + \frac{1}{2}\caE(\bar u)
  - \dot s(j) \Bigr]
\end{split}
\end{equation}
(Remember that $\dot s$ is independent of $u$ under the stationary
condition.) Restricting the set of currents even further by imposing
the condition $\caD(j) = \dot s(j)$, the above equals
\begin{equation}
  \eqref{eq: current-law} = \frac{1}{4} \bigl[ \caD(\bar j) -
  \caD(j) \bigr]
  = \frac{1}{4} \bigl[ \dot s(\bar j) - \dot s(j) \bigr]
\end{equation}
This in particular yields that the stationary current maximizes the
entropy production rate under the above two constraints, which is an
instance of the stationary maximum entropy production principle.

\subsection{Far from equilibrium}\label{noneq}

As we have seen the most remarkable feature of small fluctuations in
the close-to-equilibrium regime is that the empirical distributions
of occupations and of currents become uncorrelated. This appears to be the
fundamental reason for the entropy production principles discussed
in the previous section to be valid. An important novel feature of
the nonequilibrium statistics beyond the close-to-equilibrium regime
is that both empirical observables get coupled as we can demonstrate
via the other scaling limit introduced in~\eqref{eq: scaled}. There
we do an expansion up to leading order around the stationary density
$\rho$ and the corresponding stationary current
$\bar j(x,y) = \rho(x) \la(x,y) - \rho(y) \la(y,x)$; recall that $F$
and hence the rates $\la(x,y)$ remain fixed now. Using also the
notation
\begin{equation}\label{eq: traffic}
  \bar \tau(x,y) = \rho(x) \la(x,y) + \rho(y) \la(y,x)
\end{equation}
for the steady traffic, the scaled Lagrangian~\eqref{eq: scaled}
obtains the form:
\begin{equation}\label{eq: L-far}
\begin{split}
  L(v,\frj;F) &= \lim_{\ep\downarrow 0} \ep^{-2}
  \caL(\rho + \ep\rho\,v, \bar j + \ep \frj; F)
\\
  &= \sum_{(xy)} \frac{1}{4\bar \tau}\,\bigl[ \frj - \bar \tau  \nabla^- v
  - \bar j \nabla^+ v \bigr]^2(x,y)
\end{split}
\end{equation}
with the notation $\nabla^\pm v(x,y) = \frac{1}{2} [v(x) \pm v(y)]$.
This is the Lagrangian describing normal fluctuations around the
typical evolution, with
$\bar \tau  \nabla^- v + \bar j \nabla^+ v$ being the typical (or
expected) first-order deviation from the steady current $\bar j$. We
see that the steady traffic $\bar \tau$ plays the role of a variance
in this fluctuation law.

In the stationary regime, i.e.\ under the constraint
$\sum_{y \neq x} \frj(x,y) = 0$, the Lagrangian~\eqref{eq: L-far}
yields the rate function for the joint distribution of time-average
occupations and currents, cf.~\eqref{statfluct}. We can write it in
the form
\begin{equation}\label{eq: L-far1}
  L(v,\frj;F) = \frac{1}{2} \sum_{(xy)} \Bigl[
  \frac{1}{2\bar \tau}\, \frj^2 + \frac{\bar \tau}{2} (\nabla^- v)^2
  - \frac{\bar j}{\bar \tau}\, \frj \nabla^+ v
  + \frac{\bar j^2}{2 \bar \tau} (\nabla^+ v)^2
  \Bigr](x,y)
\end{equation}
which demonstrates that the emerged occupation-current coupling is
proportional to the stationary current and indeed vanishes only
close to equilibrium when moreover
$\bar j = O(\ep)$.

\section{Towards a more general theory}\label{noneq-full}

Adding a nonequilibrium driving not only generates nonzero steady
currents but it also modifies the steady averages of
\emph{time-symmetric} observables and their fluctuation statistics.
For a long time, the latter has not been of primary interest in
transport considerations, partially because of the success of linear
response theories in which only the currents and the entropy
production play a fundamental role. The origin of that has been discussed in Section~\ref{eq} on the structure of
close-to-equilibrium normal fluctuations in which the time-symmetric
and the time-antisymmetric sectors become totaly decoupled. Their coupling away from equilibrium, cf.~Section~\ref{noneq},  suggests that some systematic and robust description of nonequilibrium fluctuations might be achieved by
analyzing the time-antisymmetric (currents) and the time-symmetric
(e.g.\ the occupation times) observables simultaneously; this is
exactly the strategy brought up in the present paper.

\subsection{Traffic}
An important drawback of the transport theories based on stochastic
models is that we only have a direct thermodynamic interpretation
for the time-antisymmetric part of the transition rates, cf.~the
local detailed balance condition~\eqref{ldb}, whereas rather little can
\emph{generally} be said about the symmetric part and its
dependence on the nonequilibrium driving. Yet, the fluctuation
theory can help also here: instead of giving an interpretation to
the symmetric part of the rates, one can try to understand the role
of the traffic~\eqref{eq: traffic} as a time-symmetric dynamical
observable and a counterpart to the current. We have already seen
that in equilibrium the (mean) traffic coincides with
$2\ga(x,y)$, and away from equilibrium it enters, according
to~\eqref{eq: L-far1}, as a variance for normal dynamical
fluctuations. More generally and even beyond the regime of normal
fluctuations, it can be shown that the dependence of the traffic on
the driving fully determines the structure of nonequilibrium
fluctuations in the time-symmetric sector, and also specifies the
symmetric-antisymmetric coupling in a canonical way, \cite{prl,MNW}.
We give here a brief review of this approach.

The Lagrangian $\caL$ introduced in~\eqref{eq: L2-gen}--\eqref{eq:
j-explicit} has the form
\begin{equation}\label{eq: L-general}
\caL(p,j) = \sum_{x,y} p(x) [k^j(x,y)\, \psi^j(x,y)
-k^j(x,y)+\lambda(x,y)]
\end{equation}
with the modified rates $k^j$ given by~\eqref{eq: kj}: they can be
thought of as the original rates but with the modified work function
$F \rightarrow F + \psi^j$ fixed so that the current $j$ becomes
typical,
\begin{equation}\label{eq: current again}
  j(x,y) = p(x) k^j(x,y) - p(y)k^j(y,x)
\end{equation}
Therefore,
\begin{equation}\label{spil}
\caL(p,j) = \sum_{(xy)} \bigl[ j(x,y)\,\psi^j(x,y) - \tau_{p,F+\psi^j}(x,y)
+ \tau_{p,F}(x,y) \bigr]
\end{equation}
with the (mean) traffic
\[
\tau_{p,G}= p(x)\la_G(x,y) + p(y)\la_G(y,x),\qquad
\la_G(x,y) = \la_0(x,y)\,e^{\frac{\be}{2}G(x,y)}
\]
considered here as a function of the work matrix $G$; the
$\la_0(x,y) = \rho_o^{-1}(x) \ga(x,y) $
being the rates corresponding to the reversible reference dynamics
$G = 0$, cf.~\eqref{transrates}. Note that
$\la_{F + \psi^j} = k^j$ under the relation~\eqref{eq: current
again}; in fact the driving and the current appear to be conjugated
variables in the sense of a canonical
formalism, see~\cite{prl} for details.

Clearly, \eqref{spil} splits in two parts: the first term is to be
understood as an \emph{excess} of entropy production and the second
one is an excess of overall traffic. Starting from this general
scheme, we can calculate the fluctuation rate functions of arbitrary
more coarse-grained dynamical observables. If that observable is
purely time-symmetric, e.g.\ only depends on the occupations, then
the excess
$\psi^j$ is in fact a gradient, $\psi^j(x,y) = V(y) - V(x)$ for state
function $V$, and the first entropy production term in~\eqref{spil}
vanishes for currents satisfying the stationarity condition
$\sum_{y \neq x} j(x,y) = 0$.  Then, what remains is the excess traffic as
variational functional.

\subsection{Why does the entropy production govern close-to-equilibrium?}
By construction the traffic functionals are quite different from the
entropy production functionals. Yet, close to equilibrium the excess
of traffic and the excess of entropy production are related to each
other in a simple way: analogously to the scaled mean entropy production~\eqref{eq: si-E},
\begin{align}
  \caE_F(u) &= \lim_{\ep\downarrow 0} \ep^{-2}
  \si(\rho_o + \ep\rho_o u; \ep F)
\\\intertext{we consider the scaled overall traffic}
  \caT_F(u) &= \lim_{\ep\downarrow 0} \ep^{-2}
  \sum_{(xy)} \bigl[ \tau_{\rho_o + \ep\rho_o u;\,\ep F}(x,y)
  - \tau_{\rho_o + \ep\rho_o u;\, 0}(x,y) \bigr]
\end{align}
(relatively with respect to the equilibrium reference so that the
limit is well defined), and we find the relation
\begin{equation}\label{eq: ep-traffic}
  \caT_F(u) = \frac{1}{2}\, \bigl[ \caE_F(u) - \caE_0(u) \bigr]
\end{equation}
Hence, close to equilibrium the mean overall traffic is determined
by the mean entropy production. This observation remarkably
simplifies the structure of Lagrangian~\eqref{eq: L-general}, and it
finally leads to the (generalized) Onsager-Machlup theory of
Section~\ref{eq} in which the entropy production and derived
quantities are the only players.

To summarize, we understand  the dominance of entropy production and
the simple structure of the close-to-equilibrium regime as a
consequence (1) of the decoupling of the time-symmetric and the
time-antisymmetric fluctuations, (2) of the relation between the
mean traffic and the mean entropy production in this regime.

\subsection{Beyond entropy production far from equilibrium}

In a far-from-equilibrium regime the time-symmetric and the
time-antisymmetric fluctuations get coupled and the
relation~\eqref{eq: ep-traffic} is no longer valid. This gives a
motivation why it is natural to study both dynamical sectors
jointly. As we have seen, the entropy production alone is not
sufficient to describe fluctuations in either of these sectors, and
the traffic functional enters as a new important player in the
nonequilibrium fluctuation theory. We hope that more theoretical investigation and also experimental
evidence will support this line of research.\\ \\

\noindent {\bf Acknowledgment}: K.N.\ acknowledges the support
from the Grant Agency of the Czech Republic (Grant
no.~202/07/0404). C.M.\ benefits from the Belgian Interuniversity
Attraction Poles Programme P6/02. B.W.\ is an aspirant of FWO,
Flanders.\\

\bibliographystyle{plain}

\end{document}